\documentclass[aps,amssymb,twocolumn]{revtex4}
\usepackage{amssymb}
\usepackage{amsmath}
\usepackage{graphicx}
\usepackage{color}
\usepackage{textcomp}
\usepackage{ulem}
\usepackage{lipsum}
\begin{document}

\title{ $\rho$ mesons in finite magnetic field and finite temperature }
\author{Zhiyang Liu, Min Zhou, Yvming Tian, Rui Zhou, Guoyun Shao}
\author{Shijun Mao}
 \email{maoshijun@mail.xjtu.edu.cn}
\affiliation{School of Physics, Xi'an Jiaotong University, Xi'an, Shaanxi 710049, China}

\begin{abstract}
The mass spectra of $\rho$ mesons ($\rho_{Q=\pm 1}^{s_z=0,\pm 1}$ and $\rho_{Q=0}^{s_z=0,\pm 1}$) at finite magnetic field and temperature are studied in frame of the two-flavor Nambu-Jona-Lasinio model. Fully considering the breaking of translational invariance induced by external magnetic field, the analytical form of $\rho$ meson propagators have been derived in the Ritus scheme and Schwinger scheme, which gives the same algebraic formula. When solving the pole equation of $\rho$ meson propagators, multiple solutions of the meson mass appear due to the dimension reduction of their constituent quarks in magnetic fields. At vanishing temperature, we focus on the $\rho$ meson masses $M_{\rho}$ corresponding to the lowest value solution of the pole equation. $M_{\rho^{-}_+}$, $M_{\rho^{0}_+}$ and $M_{\rho^{\pm}_0}$ increase with magnetic field. $M_{\rho^{+}_+}$ firstly decreases and then becomes saturated with increasing magnetic field. $M_{\rho^0_0}$ is not sensitive to magnetic field. These results are consistent with the available LQCD simulations. At finite temperature, we discuss the lowest four/five solutions of $\rho$ meson masses $M^{i=0,1,2,3,4}_{\rho}$. With fixed magnetic field, they decrease with temperature, and approach the mass sum of their constituent quarks at high temperature. The mass solution $M^{i}_{\rho}$ for different mesons $\rho_+^{0,\pm}$ and $\rho_0^{0,\pm}$ may become degenerate at finite magnetic field and temperature.
\end{abstract}

\date{\today}
\maketitle

\section{Introduction}
\label{intro}
Strong electromagnetic fields exist in Quantum Chromodynamics (QCD) systems like high energy nuclear collisions~\cite{kharzeev2008effects,kharzeev2008chiral,kharzeev2009chern,kharzeev2009chiral,skokov2009estimate,voronyuk2011electromagnetic,deng2012event,bacsar2012conformal}, compact stars~\cite{kouveliotou1998x,ferrer2010equation,paulucci2011equation,isayev2011finite,ferrer2013magnetism,mallick2014deformation,menezes2014repulsive,chu2015quark,cao2015effects} and early universe~\cite{vachaspati1991magnetic,enqvist1993primordial,cheng1994primordial,baym1996magnetic}. The electromagnetic interaction provides a sensitive probe for the hadron structure. At hadron level, without considering the inner structure, neutral hadrons are blind to electromagnetic fields and their properties in hot and dense medium are not directly affected by the fields. At quark level, however, the electromagnetic interaction of the charged constituent quarks leads to a sensitive dependence of the neutral mesons (such as $\pi_0,\ K_0, \ \eta, \ \rho_0,\ \phi$) on the external electromagnetic fields~\cite{ayala2021magnetic,aguirre2017modification,ayala2018magnetic,das2020neutral,nasser2019chiral,luschevskaya2015magnetic,ding2021chiral,avancini2017pi0,avancini2016properties,fayazbakhsh2012properties,fayazbakhsh2013weak,sheng2021pole,sheng2022impacts,gomez2018neutral,avancini2019neutral,andersen2012thermal,luschevskaya2016magnetic,wang2018meson,zhang2016properties,liu2018neutral,li2021gauge,li2022light,li2023inverse,coppola2019neutral,xu2021effect,orlovsky2013nambu,hattori2016mesons,simonov2016pion,kojo2021neutral,mishra2021strange,avancini2021light,Cha,larina2014rho,Wei,Li:2025wqb,Mei:2022dkd,mao2017effect,shengxinli,ghosh2020thermo,avancini2022magnetized}. For charged meson study at quark level, (such as $\pi_\pm,\ K_\pm,\ \rho_\pm$), the complexity lies in the lack of translational invariance in external electromagnetic fields, and the Fourier-like transformation between coordinate and momentum spaces should be carried out by the eigenfunctions of the magnetized field equation for mesons~\cite{andersen2012thermal,luschevskaya2016magnetic,wang2018meson,zhang2016properties,liu2018neutral,li2021gauge,li2022light,li2023inverse,coppola2019neutral,xu2021effect,orlovsky2013nambu,hattori2016mesons,simonov2016pion,kojo2021neutral,mishra2021strange,avancini2021light,Cha,larina2014rho,colucci2014chiral,hidaka2013charged,coppola2018charged,mao2019pions,Tian,chao,luschevskaya2017determination,bali2018meson,carlomagno2022charged,Sccoccola,gomez2023charged,Mei:2026xlj}.

When the magnetic field is of strength comparable with the strong interaction energy scale, such as $eB\sim m^2_\pi$, the quark structure of hadrons should be taken into account. We make use of the two flavor Nambu-Jona-Lasinio (NJL) model~\cite{klevansky1992nambu,nambu1961dynamical,volkov1992effective,hatsuda1994qcd,buballa2005njl} to study the mass spectra of $\rho$ mesons ($\rho_{Q=\pm 1}^{s_z=0,\pm 1}$ and $\rho_{Q=0}^{s_z=0,\pm 1}$) under magnetic field. Different from the previous work~\cite{ghosh2020thermo,avancini2022magnetized,carlomagno2022charged,Sccoccola,gomez2023charged}, we conduct the analytical derivations for $\rho$ meson propagators by using Ritus scheme and Schwinger scheme of particle propagators in the magnetic field, and prove that different schemes lead to the same algebraic formula. In addition, we numerically solve the pole mass of $\rho$ mesons at finite magnetic field and temperature. Due to the dimension reduction of their constituent quarks in magnetic fields, multiple solutions appear. The degeneracy for $\rho$ meson mass spectra is analyzed.

The rest of paper is arranged as follows. We introduce the magnetized $SU(2)$ NJL model, and derive the analytical formula for meson propagators in Ritus scheme and Schwinger scheme of particle propagators in Sec.\ref{form}, together with Appendix. The numerical results and analysis of meson masses at finite magnetic field and temperature are presented in Sec.\ref{numerical}. The summary and outlook is written in Sec.\ref{sum}.

\section{Framework}
\label{form}
We perform our study in the framework of two-flavor Nambu-Jona-Lasinio(NJL) model, which is presented by the Lagrangian density in terms of quark fields $\psi$~\cite{klevansky1992nambu,nambu1961dynamical,volkov1992effective,hatsuda1994qcd,buballa2005njl}
\begin{align}
		\label{njl}
		{\cal L} =& \bar{\psi}\left(i \gamma^{\mu} {D_{\mu}}-m_{0} \right) \psi +G_S\left[(\bar{\psi} \psi)^{2}+\left(\bar{\psi} i \gamma^{5} \vec{\tau} \psi\right)^{2}\right]\nonumber\\
		&-G_V\left[(\bar{\psi} \gamma^{\mu} \tau_{a} \psi)^{2}+\left(\bar{\psi} \gamma^{\mu} \gamma^{5} \tau_{a} \psi\right)^{2}\right].
\end{align}
The covariant derivative $D_{\mu} = \partial_{\mu} - iq_{f}A_{\mu}$ is defined in the magnetic field. We consider the external magnetic field $\mathbf{B} = (0,0,B)$ along the $z$-axis, with $A_{\mu} = (0,0,Bx_{1},0)$ in the Landau gauge. $q_{f} = \mathrm{diag}(Q_{u}, Q_{d}) = (2e/3, -e/3)$ is quark electric charge matrix in flavor space. We assume the same current mass for $u$ and $d$ quark $m_{0} = m_{u} = m_{d}$. $\vec{\tau} = (\tau_{1}, \tau_{2}, \tau_{3})$ are the Pauli matrices in flavor space. $G_{S}$ and $G_{V}$ are the coupling constants for the scalar and vector channels, respectively.

%
%
%
%

In the NJL model, the mesons are regarded as quantum fluctuations or collective modes. With the random phase approximation (RPA) method~\cite{klevansky1992nambu,nambu1961dynamical,volkov1992effective,hatsuda1994qcd,buballa2005njl,zhuang1994j}, the meson propagator can be expressed as a sum of infinite-order quark loop diagrams. Through the Dyson-Schwinger equation, we can relate the $\rho$ meson propagator to the one-loop polarization function
\small
\begin{align}
	\label{DS}
	D^{\mu \nu}_{M^\prime}(x,z)&=-2G_Vg^{\mu \nu}\delta(x-z)\\
	&\ \ -\int d^4y2G_Vg^{\mu \alpha}\Pi_{{M^\prime},\alpha \beta}(x,y)D^{\beta \nu}_{M^\prime}(y,z),\nonumber
\end{align}
where the one-loop polarization function
\small
\begin{equation}
	\label{jihua}
	\Pi_{{M^\prime}}^{\mu\nu}(x,z)=i\mathrm{Tr} \left[ \Gamma^{*^{\mu}}_{M^\prime} S(x,y)\Gamma_{M^\prime}^{\nu} S(y,z)\right]
\end{equation}
is constructed from quark propagator $S(x,y)=\mathrm{diag}(S_{u}, S_{d})$ in flavor space (see Eq.\eqref{sfritus} or Eq.\eqref{QkSs}) and vertices $\Gamma_{{M^\prime}}$
\begin{equation}
	\begin{split}
		\Gamma_{{M^\prime}}^{\nu}=\left\{\begin{array}{ll}
			\tau_{+} \gamma^{\nu} & {M^\prime}=\rho_{+} \\
			\tau_{-} \gamma^{\nu} & {M^\prime}=\rho_{-} \\
			\tau_{3} \gamma^{\nu} & {M^\prime}=\rho_{0},
		\end{array}\right. \\
		\Gamma_{{M^\prime}}^{*\mu}=\left\{\begin{array}{ll}
			\tau_{-} \gamma^{\mu} & {M^\prime}=\rho_{+} \\
			\tau_{+} \gamma^{\mu} & {M^\prime}=\rho_{-} \\
			\tau_{3} \gamma^{\mu} & {M^\prime}=\rho_{0}.
		\end{array}\right.
	\end{split}
\end{equation}
%

Note that $\rho_{+}$ and $\rho_{-}$ mesons have symmetric form, and they are degenerate in the mass spectra at finite magnetic field and temperature. In the following discussion, we only take $\rho_{+}$ meson as an example of the charged $\rho$ mesons.
The presence of an external magnetic field leads to the off-diagonal term in the polarization function tensor of the $\rho_{+}$ meson. By diagonalizing Eq~\eqref{DS}, we obtain the polarization function and propagator for $\rho_{+}$ mesons with non-zero spin
\begin{align}
	\label{gouxing}
	\Pi_{\rho_+}(s_z=\pm1)&=\frac{\Pi_{\rho_+}^{11}+\Pi_{\rho_+}^{22}\pm i(\Pi_{\rho_+}^{12}-\Pi_{\rho_+}^{21})}{2}, \\
	D_{\rho_+}(s_z=\pm1)&=\frac{D_{\rho_+}^{11}+D_{\rho_+}^{22}\pm i(D_{\rho_+}^{12}-D_{\rho_+}^{21})}{2}.
\end{align}
The polarization function and propagator for $\rho_{+}$ meson with zero spin are
\begin{align}
	\Pi_{\rho_+}(s_z=0)&=\Pi_{\rho_+}^{33},\\
   D_{\rho_+}(s_z=0)&=D_{\rho_+}^{33}.
\end{align}
For $\rho_0$ mesons, we have
\begin{align}
\Pi_{\rho_0}(s_z=\pm1)&=\frac{\Pi_{\rho_0}^{11}+\Pi_{\rho_0}^{22}}{2},\\
	D_{\rho_0}(s_z=\pm1)&=\frac{D_{\rho_0}^{11}+D_{\rho_0}^{22}}{2},\\
\Pi_{\rho_0}(s_z=0)&=\Pi_{\rho_0}^{33},\\
D_{\rho_0}(s_z=0)&=D_{\rho_0}^{33},
\end{align}
with degenerate $\rho_0^\pm$ mesons. Hence, Eq.~\eqref{DS} for the $\rho$ mesons ($\rho_{Q=\pm 1}^{s_z=0,\pm 1}$ and $\rho_{Q=0}^{s_z=0,\pm 1}$) can be rewritten as
\begin{eqnarray}
	\label{DS1}
	\small{D_{M}(x,z)=2G_V\delta(x-z)+\int d^4y2G_V\Pi_{M}(x,y)D_{M}(y,z)},
\end{eqnarray}
where $M$ represents $\rho_{Q=\pm 1}^{s_z=0,\pm 1}$ and $\rho_{Q=0}^{s_z=0,\pm 1}$ mesons. In the following, we use the notation $\rho_{Q=\pm 1}^{s_z=0,\pm 1}$ and $\rho_{Q=0}^{s_z=0,\pm 1}$ for $\rho$ mesons with different electric charges $Q$ and spin $s_z$.

To investigate the mass spectra of $\rho$ mesons, we need to derive the $\rho$ meson propagator in (conserved) momentum space.
For neutral $\rho_0$ mesons, the conserved momentum in magnetic field is the usually defined momentum $\bar{k}=(k_0,k_1,k_2,k_3)$. The Fourier transformation from coordinate space to momentum space is carried by the plane wave $e^{-i\bar{k} x}$, which is the eigenfunction of the $\rho_0$ meson field equation in magnetic field (magnetized Proca equation). $\rho_0$ meson polarization function and propagator in momentum space can be derived through
\begin{align}
	\label{zxbiaoxiang}
	\Pi_{M}(\bar{k})=\int d^4(x-y) e^{i\bar{k}(x-y)}\Pi_{M}(x,y),\\
\label{zxbiaoxiang1}
	D_{M}(\bar{k})=\int d^4(x-y) e^{i\bar{k}(x-y)}D_{M}(x,y).
\end{align}

For charged $\rho_+$ mesons, the conserved momentum in magnetic field is $\bar{k}=(k_0,0,-s_M\sqrt{(2n-2s_M s_z+1)|Q_MB|},k_3)$. The corresponding eigenfunction of the $\rho_+$ meson field equation in magnetic field (magnetized Proca equation) is
\begin{eqnarray}
	\label{jie}
	F_{n-s_Ms_z}(x,k)&=&e^{i(k_0x_0+k_2x_2+k_3x_3)}g_{n}^{s_M}(x_1,k_2),\\
g_n^{s_M}(x_1,k_2)&=&\phi(x_1/l_M-s_M k_2 l_M),\\
	\phi(z)&=&(2^n n!\sqrt{\pi}l_M)^{-1/2}e^{-z^2/2}H_n(z),
\end{eqnarray}
with the Landau level $n$, $s_M = \mathrm{sign}(Q_MB)$, $l_M^2=|Q_MB|^{-1}$ and $s_z = (0,\pm 1)$.

The Fourier-like transformation from coordinate space to momentum space for charged $\rho_+$ mesons is carried by the eigenfunction $F_{n-s_Ms_z}(x,k)$, and the $\rho_+$ propagator and polarization function in momentum space are
\begin{align}
	\label{biaoxiang}
	\Pi_{M}(\bar{k})=\int d^4x d^4y F^*_n(x,k)\Pi_{M}(x,y)F_n(y,k),\\
\label{biaoxiang1}
	D_{M}(\bar{k})=\int d^4x d^4y F^*_n(x,k)D_{M}(x,y)F_n(y,k).
\end{align}

Substitute Eq~\eqref{zxbiaoxiang} and Eq~\eqref{zxbiaoxiang1} (Eq~\eqref{biaoxiang} and Eq~\eqref{biaoxiang1}) into Eq~\eqref{DS}, propagators of neutral $\rho_{Q=\pm 1}^{s_z=0,\pm 1}$ mesons (charged $\rho_{Q=0}^{s_z=0,\pm 1}$ mesons) in momentum space can be written as
\begin{equation}
\label{prorho}
	D_M(\bar{k})=\frac{2G_V}{1-2G_V\Pi_M(\bar{k})},
\end{equation}
and the masses of $\rho$ mesons are determined by the pole equations with the lowest momentum
\begin{equation}
\label{poleeq}
	1-2G_V\Pi_M(\bar{k})=0.
\end{equation}

In order to give the analytical form of the polarization function in Eq.\eqref{poleeq} for $\rho_{Q=\pm 1}^{s_z=0,\pm 1}$ and $\rho_{Q=0}^{s_z=0,\pm 1}$ mesons, we define the notations $J_1$ and $J_2(k_0)$
\begin{align}
	J_1&=i\int\frac{dp_0}{2\pi}\left[\frac{1}{(p_0+k_0)^2-E_{n'}^2}+\frac{1}{p_0^2-E_n^2}\right]\\
	&=-\frac{1}{2}\left[\frac{tanh(\frac{E_{n'}}{2T})}{E_{n'}}+\frac{tanh(\frac{E_{n}}{2T})}{E_{n}}\right],\nonumber\\
	J_2(k_0)&=i\int\frac{dp_0}{2\pi}\frac{1}{\left[(p_0+k_0)^2-E_{n'}^2\right](p_0^2-E_n^2)}\\
	&=-\frac{1}{4E_{n'}E_n}\bigg[\frac{f(E_n)-f(E_{n'})}{k_0-E_{n'}+E_n}+\frac{f(-E_{n'})-f(E_n)}{k_0+E_{n'}+E_n}\nonumber\\
	&+\frac{f(E_{n'})-f(-E_n)}{k_0-E_{n'}-E_n}+\frac{f(-E_n)-f(-E_{n'})}{k_0+E_{n'}-E_n}\bigg].\nonumber
\end{align}

For $\rho_{Q=0}^{s_z=0,\pm 1}$ mesons, the composite quarks are $u$ and $\bar u$ ($d$ and $\bar d$), and thus the polarization functions with vanishing momentum ${\vec k}={\vec 0}$ can be derived as
\begin{align}
	\label{rho00}
	&\Pi \rho _{0}^{0}(k_0)=N_{c}\sum\limits _{f,n,n'}\frac{|Q_{f} B|}{2\pi }\int \frac{dq_{3}}{2\pi }\\
	&\ \left[J_1-(k_0^2-4q_3^2)J_2(k_0)\right]\left(\delta _{n',n} +\delta _{n'-1,n-1}\right),\nonumber\\
	\label{rho01}
	&\Pi \rho _{0}^{\pm}(k_0)=N_{c}\sum\limits _{f,n,n'}\frac{|Q_{f} B|}{2\pi }\int \frac{dq_{3}}{2\pi }\\
	&\ \left[J_1-(k_0^2-2(n+n')|Q_fB|)J_2(k_0)\right]\left(\delta _{n'-1,n} +\delta _{n',n-1}\right),\nonumber
\end{align}
with quark energy $E_{n'}=\sqrt{2n'|Q_fB|+q_3^2+m_q^2}$, $E_{n}=\sqrt{2n|Q_fB|+q_3^2+m_q^2}$ in $J_1$ and $J_2(k_0)$, and summation over flavor $f=u,d$.

For $\rho_{Q=+ 1}^{s_z=0,\pm 1}$ mesons, the composite quarks are $u$ and $\bar d$, and thus the polarization functions with $n=0,\ k_3=0$ can be derived as
\begin{align}
	\label{rho11}
	&\Pi{\rho _{+}^{+}}(k_{0})=4N_{c}\sum\limits _{n,n'}Z(n,n')\int \frac{dq_{3}}{( 2\pi )^{2}}\\
	&\ \left[J_1-(k_0^2-2n'|Q_uB|-2n|Q_dB|)J_2(k_0)\right],\nonumber \\
	\label{rho12}
	&\Pi{\rho _{+}^{-}}(k_{0})=4N_{c}\sum\limits _{n,n'}I_nI_{n'}Z(n-1,n'-1) \int \frac{dq_{3}}{( 2\pi )^{2}}\\
	&\ \left[J_1-(k_0^2-2n'|Q_uB|-2n|Q_dB|)J_2(k_0)\right],\nonumber \\
	\label{rho13}
	&\Pi{\rho _{+}^{0}}(k_{0})=2N_{c}\sum\limits _{n,n'}I_nZ(n-1,n') \int \frac{dq_{3}}{( 2\pi )^{2}}\\
	&\ \left[J_1-(k_0^2+(4I_{n'}-1)2n'|Q_uB|-2n|Q_dB|-4q_3^2)J_2(k_0)\right] \nonumber\\
	&\ +2N_c\sum\limits _{n,n'}I_{n'}Z(n,n'-1) \int \frac{dq_{3}}{( 2\pi )^{2}}\nonumber\\
	&\ \left[J_1-(k_0^2+(4I_n-1)2n|Q_dB|-2n'|Q_uB|-4q_3^2)J_2(k_0)\right],\nonumber
\end{align}
with $u$ quark energy $E_{n'}=\sqrt{2n'|Q_uB|+q_3^2+m_q^2}$ and $d$ quark energy $E_{n}=\sqrt{2n|Q_dB|+q_3^2+m_q^2}$ in $J_1$ and $J_2(k_0)$, and $I_n=1-\delta_{n,0}$. The factor $Z(n,n')$ is a negative binomial distribution
\begin{align}
		&Z(n,n')=\frac{1}{3}\times \frac{2}{3}|eB|C_{n+n'}^{n}\left(\frac{2}{3}\right)^n\left(\frac{1}{3}\right)^{n'}, \\
&C_{n+n'}^{n}=\frac{(n+n')!}{n!n'!}, \nonumber\\
    &{\text {with}}\  \sum_nZ(n,n') =|Q_uB|,\ {\text {and}}\ \sum_{n'}Z(n,n')=|Q_dB|, \nonumber
\end{align}
which is different from the Kronecker delta function $\delta_{n,n'}$ in neutral $\rho_0$ mesons (see Eq\eqref{rho00} and Eq\eqref{rho01}).

The main result in this section is to derive the meson propagator (see Eq.\eqref{prorho}, Eq.\eqref{rho00}, Eq.\eqref{rho01}, Eq.\eqref{rho11}, Eq.\eqref{rho12} and Eq.\eqref{rho13}) by using the quark propagator (see Eq.\eqref{sfritus} or Eq.\eqref{QkSs}). Note that the effective quark mass $m_q$ is determined through gap equation
\begin{equation}
	m_q(1-2G_SJ_0)-m_0=0,
\label{mqeb}
\end{equation}
\begin{equation}
	J_0=N_c\sum_{f,n}\alpha_n\frac{|Q_fB|}{2\pi}\int\frac{dq_3}{2\pi}\frac{\tanh{(\frac{E_f}{2T})}}{E_f},
\end{equation}
with $E_{f} = \sqrt{m_{q}^{2} + 2n|Q_fB|+q_3^2}$ and $\alpha_n=2-\delta_{n,0}$.

There are two equivalent ways to treat the particle propagators in magnetic field, the Schwinger scheme~\cite{andersen2016phase,miransky2015quantum} and the Ritus scheme~\cite{ritus1972radiative,leung2006gauge,elizalde2002neutrino}. Within the Ritus scheme and the Schwinger scheme, we obtain same algebraic formula for the polarization function in Eq.\eqref{rho00}, Eq.\eqref{rho01}, Eq.\eqref{rho11}, Eq.\eqref{rho12} and Eq.\eqref{rho13}. In the following part, we give the derivations and proof.


\subsection{Ritus Scheme}
In Ritus scheme, one can well-define the Fourier-like transformation for the particle propagator from the conserved Ritus momentum space to coordinate space~\cite{elizalde2002neutrino,ritus1972radiative,leung2006gauge}. The quark propagator with flavor $f$ in coordinate space can be expressed as
\begin{equation}
\label{sfritus}
	S_f(x,y)=\sum_{n}\int\frac{d^3p}{(2\pi)^3}e^{ip(x-y)}P_n(x_1,p_2)\mathcal{D}_f(\bar{p})P_n(y_1,p_2),
\end{equation}
where $p = (p_0, 0, p_2, p_3)$ is the Fourier momentum, $\bar{p} = (p_0, 0, -s_f \sqrt{2n|Q_f B|}, p_3)$ is the conserved momentum in Ritus space, $n$ is the Landau level, $s_f = \mathrm{sign}(Q_f B)$, and $\mathcal{D}_f(\bar{p}) = (\gamma \cdot \bar{p} - m_q)^{-1}$. The projection operator in the Ritus scheme is
\begin{align}
	P_n(x_1,p_2)&=\frac{1}{2}\left[ g_n^{s_f}(x_1,p_2)+I_ng_{n-1}^{s_f}(x_1,p_2)\right]\nonumber\\
	&+\frac{is_f}{2}\left[ g_n^{s_f}(x_1,p_2)-I_ng_{n-1}^{s_f}(x_1,p_2)\right]\gamma^1\gamma^2.
\end{align}

With Eq~\eqref{jihua} and Eq~\eqref{sfritus}, the polarization function of the $\rho$ mesons can be expressed as
\begin{align}
	\label{Rjihua}
	&\Pi_{M}^{\mu\nu}(x,y)=2iN_{c}\sum\limits _{n,n^{'}}\int \frac{d^{3}p d^{3}q}{( 2\pi )^{6}} e^{i \left(p -q \right)( x-y)}\times \nonumber\\
	 & \mathrm{Tr}_{D}\left[P_{n}( x_{1} ,q_{2}) \gamma ^{\mu} P_{n'}( x_{1} ,p_{2})\frac{\gamma \cdot \overline{p} +m_{q}}{\overline{p}^{2} -m_{q}^{2}}\right.\nonumber\\
	&\ \ \times \left. P_{n'}( y_{1} ,p_{2}) \gamma ^{\nu} P_{n}( y_{1} ,q_{2})\frac{\gamma \cdot \overline{q} +m_{q}}{\overline{q}^{2} -m_{q}^{2}}\right].
\end{align}
For neutral $\rho_0$ mesons, the polarization function can be written as
\begin{align}
	\label{rho11jihua3}
	&\textstyle\Pi {\rho _{0}^{\pm}}(x,y)=2iN_c\sum\limits _{f,n,n^{'}}\int \frac{d^{3}p d^{3}q}{( 2\pi )^{6}}e^{i\left(p -q\right)( x-y)}\\
	&\textstyle\times\frac{\left(\alpha_-^{s_f}\alpha_+^{s_f}+\alpha_+^{s_f}\alpha_-^{s_f}\right)\left(\bar{p}_{0}\bar{q}_{0} -\bar{p}_{3}\bar{q}_{3} -m_{q}^{2}\right)}{\left(\bar{p}^{2} -m_{q}^{2}\right)\left(\bar{q}^{2} -m_{q}^{2}\right)},\nonumber\\
	\label{rho11jihua4}
	&\textstyle\Pi {\rho _{0}^{0}}(x,y)=2iN_c\sum\limits _{f,n,n^{'}}\int \frac{d^{3}p d^{3}q}{( 2\pi )^{6}}e^{i\left(p -q\right)( x-y)}\\
	&\textstyle\times\left[\frac{\left(\alpha_+^{s_f}\alpha_+^{s_f}+\alpha_-^{s_f}\alpha_-^{s_f}\right)\left(\overline{p}_{0}\overline{q}_{0} +\overline{p}_{3}\overline{q}_{3} -m_{q}^{2}\right)}{\left(\overline{p}^{2} -m_{q}^{2}\right)\left(\overline{q}^{2} -m_{q}^{2}\right)}-\frac{\left(\beta_+^{s_f}\beta_+^{s_f}+\beta_-^{s_f}\beta_-^{s_f}\right)\overline{p}_{2}\overline{q}_{2}}{\left(\overline{p}^{2} -m_{q}^{2}\right)\left(\overline{q}^{2} -m_{q}^{2}\right)}\right], \nonumber
\end{align}
with
\begin{eqnarray}
	\label{ABC}
	&\alpha_+^{s_f}(k_2)&=g_n^{s_f}(x_1,k_2)g_n^{s_f}(y_1,k_2),\\
\label{A2}
	&\alpha_-^{s_f}(k_2)&=I_ng_{n-1}^{s_f}(x_1,k_2)I_ng_{n-1}^{s_f}(y_1,k_2),\\
\label{B1}
	&\beta_+^{s_f}(k_2)&=g_{n}^{s_f}(x_1,k_2)I_ng_{n-1}^{s_f}(y_1,k_2),\\
\label{B2}
	&\beta_-^{s_f}(k_2)&=I_ng_{n-1}^{s_f}(x_1,k_2)g_{n}^{s_f}(y_1,k_2).
\end{eqnarray}
For charged $\rho_+$ mesons, we have the polarization function
\begin{align}
	\label{rho11jihua}
	&\textstyle\Pi{\rho_+^{+}}(x,y)=8iN_{c}\sum\limits _{n,n^{'}}\int \frac{d^{3}p d^{3}q}{( 2\pi )^{6}} e^{i\left(p -q\right)( x-y)}\\
	&\textstyle\times\frac{ \alpha_+^{s_u}\alpha_+^{s_d}\left(\overline{p}_{0}\overline{q}_{0} -\overline{p}_{3}\overline{q}_{3} -m_{q}^{2}\right)}{\left(\overline{p}^{2} -m_{q}^{2}\right)\left(\overline{q}^{2} -m_{q}^{2}\right)},\nonumber\\
	\label{rho11jihua1}
	&\textstyle\Pi{\rho_+^{-}}(x,y)=8iN_{c}\sum\limits _{n,n^{'}}\int \frac{d^{3}p d^{3}q}{( 2\pi )^{6}} e^{i\left(p -q\right)( x-y)} \\
	&\textstyle\times\frac{\alpha_-^{s_u}\alpha_-^{s_d}\left(\overline{p}_{0}\overline{q}_{0} -\overline{p}_{3}\overline{q}_{3} -m_{q}^{2}\right)}{\left(\overline{p}^{2} -m_{q}^{2}\right)\left(\overline{q}^{2} -m_{q}^{2}\right)},\nonumber\\
	\label{rho11jihua2}
	&\textstyle\Pi {\rho _{+}^{0}}(x,y)=4iN_{c}\sum\limits _{n,n^{'}}\int \frac{d^{3}p d^{3}q}{( 2\pi )^{6}} e^{i\left(p -q\right)( x-y)}\\
	&\textstyle\times\left[ \frac{\left(\alpha_+^{s_u}\alpha_-^{s_d}+\alpha_-^{s_u}\alpha_+^{s_d}\right)\left(\overline{p}_{0}\overline{q}_{0} +\overline{p}_{3}\overline{q}_{3} -m_{q}^{2}\right)}{\left(\overline{p}^{2} -m_{q}^{2}\right)\left(\overline{q}^{2} -m_{q}^{2}\right)} -\frac{\left(\beta_+^{s_u}\beta_-^{s_d}+\beta_-^{s_u}\beta_+^{s_d}\right)\overline{p}_{2}\overline{q}_{2}}{\left(\overline{p}^{2} -m_{q}^{2}\right)\left(\overline{q}^{2} -m_{q}^{2}\right)}\right].\nonumber
\end{align}

Making use of Eq.\eqref{flnn} and Eq.\eqref{flnn1} in appendix, it can be shown that polarization function in Ritus scheme (see Eq.\eqref{rho11jihua3}, Eq.\eqref{rho11jihua4}, Eq.\eqref{rho11jihua}, Eq.\eqref{rho11jihua1} and Eq.\eqref{rho11jihua2}) can be rewritten in the form of Schwinger scheme (see Eq.\eqref{rho11jihua3new}, Eq.\eqref{rho11jihua4new}, Eq.\eqref{rho11jihuanew}, Eq.\eqref{rho11jihua1new} and Eq.\eqref{rho11jihua2new}). In this way, we establish the mathematical equivalence between the Schwinger and Ritus schemes.

\subsection{Schwinger Scheme}
In Schwinger scheme~\cite{andersen2016phase,miransky2015quantum}, the particle propagator in magnetic field is composed of two parts, one is related to the Schwinger phase which breaks the translational invariance, and the other is a Fourier transformation of the translational invariant propagator. Quark propagator of flavor $f$ in the Schwinger scheme reads
\begin{equation}
	\label{QkSs}
	S_f(x,y)=e^{i\Phi_f(x_\perp,y_\perp)}\int\frac{d^4p}{(2\pi)^4}e^{ip(x-y)}\tilde{S}_f(p_\perp,p_\parallel).
\end{equation}
Here, the Schwinger phase $\Phi_f(x_\perp,y_\perp)=s_f (x + y)(x - y) / 2l_f^2$ represents the breaking of translation invariance under external magnetic field. $\tilde{S}_f(p_\perp, p_\parallel)$ is translational invariant and can be expressed in the Landau level form,
\begin{align}
	\label{tildeS}
	\tilde{S}_{f}( p_{\perp } ,p_{\parallel }) =&e^{-p_{\perp }^{2} l_{f}^{2}} \sum\limits _{n}\frac{( -1)^{n} D_{n}( p_{\perp } ,p_{\parallel })}{p_{\parallel }^{2} -m_{q}^{2} -2n|Q_{f} B|},\\
	\label{Dn}
	D_{n}( p_{\perp } ,p_{\parallel }) =&4( \gamma _{\perp } \cdotp p_{\perp }) L_{n-1}^{1}\left( 2p_{\perp }^{2} l_{f}^{2}\right) +\left( \gamma ^{0} p_{0} -\gamma ^{3} p_{3} +m_q\right)\nonumber\\
	&\left[\mathcal{P}_+ L_{n}\left( 2p_{\perp }^{2} l_{f}^{2}\right) - \mathcal{P}_- L_{n-1}\left( 2p_{\perp }^{2} l_{f}^{2}\right)\right],
\end{align}
where $\mathcal{P}_\pm=(1\pm i\gamma^1\gamma^2s_f)$ is the projection operator with $s_f = \mathrm{sign}(Q_f B)$, and $L_n(z), L_n^a(z)$ are the Laguerre polynomials and generalized Laguerre polynomials, respectively.

Substituting the quark propagator in Eq.\eqref{QkSs} into Eq.\eqref{jihua}, we have the polarization function with straightforward calculations
\begin{align}
	\label{rho11jihua3new}
	&\textstyle\Pi {\rho _{0}^{\pm}}(x,y)=-4iN_c\sum\limits _{f,n,n^{'}}\int \frac{d^{4}\tilde{p} d^{4}\tilde{q}}{( 2\pi )^{8}} e^{i\left(\tilde{p} -\tilde{q}\right)(x-y)}(-1)^{n+n'}\\
	&\textstyle\times e^{-\tilde{p}_\perp^2l_f^2}e^{-\tilde{q}_\perp^2l_f^2}\frac{\mathcal{N}(l_f,l_f)(\bar{p}_0\bar{q}_0-\bar{p}_3\bar{q}_3-m_q^2)}{(\bar{p}^2-m_q^2)(\bar{q}^2-m_q^2)},\nonumber\\
	\label{rho11jihua4new}
	&\textstyle\Pi {\rho _{0}^{0}}(x,y)=4iN_c \sum\limits _{f,n,n'}\int \frac{d^{4}\tilde{p} d^{4}\tilde{q}}{( 2\pi )^{8}} e^{i\left(\tilde{p} -\tilde{q}\right)(x-y)}(-1)^{n+n'}\\
	&\textstyle\times e^{-\tilde{p}_\perp^2l_f^2}e^{-\tilde{q}_\perp^2l_f^2}\frac{\left[\mathcal{J}(l_f,l_f)+\mathcal{K}(l_f,l_f)\right](\bar{p}_0\bar{q}_0+\bar{p}_3\bar{q}_3-m_q^2)-\mathcal{M}(l_f,l_f)(\tilde{p}_\perp\cdot\tilde{q}_\perp)}{(\bar{p}^2-m_q^2)(\bar{q}^2-m_q^2)},\nonumber\\
	\label{rho11jihuanew}
	&\textstyle\Pi{\rho_+^{+}}(x,y)=8iN_c e^{i\Phi(x_\perp,y_\perp)}\sum\limits _{n,n'}\int \frac{d^{4}\tilde{p} d^{4}\tilde{q}}{( 2\pi )^{8}} e^{i\left(\tilde{p} -\tilde{q}\right)(x-y)}(-1)^{n+n'} \nonumber\\
	&\textstyle\times e^{-\tilde{p}_\perp^2l_u^2}e^{-\tilde{q}_\perp^2l_d^2}\frac{2\mathcal{J}(l_u,l_d)(\bar{p}_0\bar{q}_0-\bar{p}_3\bar{q}_3-m_q^2)}{(\bar{p}^2-m_q^2)(\bar{q}^2-m_q^2)},\\
	\label{rho11jihua1new}
	&\textstyle\Pi{\rho_+^{-}}(x,y)=8iN_c e^{i\Phi(x_\perp,y_\perp)}\sum\limits _{n,n'}\int \frac{d^{4}\tilde{p} d^{4}\tilde{q}}{( 2\pi )^{8}} e^{i\left(\tilde{p} -\tilde{q}\right)(x-y)}(-1)^{n+n'}\nonumber\\
	&\textstyle\times e^{-\tilde{p}_\perp^2l_u^2}e^{-\tilde{q}_\perp^2l_d^2}\frac{2\mathcal{K}(l_u,l_d)(\bar{p}_0\bar{q}_0-\bar{p}_3\bar{q}_3-m_q^2)}{(\bar{p}^2-m_q^2)(\bar{q}^2-m_q^2)}, \\
	\label{rho11jihua2new}
	&\textstyle\Pi {\rho _{+}^{0}}(x,y)=-8iN_c e^{i\Phi(x_\perp,y_\perp)}\sum\limits _{n,n'}\int \frac{d^{4}\tilde{p} d^{4}\tilde{q}}{( 2\pi )^{8}} e^{i\left(\tilde{p} -\tilde{q}\right)(x-y)}(-1)^{n+n'} \nonumber\\
	&\textstyle\times e^{-\tilde{p}_\perp^2l_u^2}e^{-\tilde{q}_\perp^2l_d^2}\frac{\mathcal{N}(l_u,l_d)(\bar{p}_0\bar{q}_0+\bar{p}_3\bar{q}_3-m_q^2)+\mathcal{M}(l_u,l_d)(\tilde{p}_\perp\cdot\tilde{q}_\perp)}{(\bar{p}^2-m_q^2)(\bar{q}^2-m_q^2)},
\end{align}
and
\begin{align}
	\Phi_{M}(x_{\perp},y_{\perp}) &=s_M (x + y)(x - y) / 2l_M^2, \nonumber\\
	\mathcal{J}(l_{f},l_{f'})&=2L_{n'}(2p_\perp^2l_f^2)L_n(2q_\perp^2l_{f'}^2),\nonumber\\
	\mathcal{K}(l_{f},l_{f'})&=2I_nI_{n'}L_{n'-1}(2p_\perp^2l_f^2)L_{n-1}(2q_\perp^2l_{f'}^2),\nonumber\\
	\mathcal{M}(l_{f},l_{f'})&=16I_{n'}I_nL_{n'-1}^1(2p_\perp^2l_f^2)L_{n-1}^1(2q_\perp^2l_{f'}^2),\nonumber\\
	\mathcal{N}(l_{f},l_{f'})&=2I_nL_{n'}(2p_\perp^2l_f^2)L_{n-1}(2q_\perp^2l_{f'}^2),\nonumber\\
	&+2I_{n'}L_{n'-1}(2p_\perp^2l_f^2)L_{n}(2q_\perp^2l_{f'}^2).\nonumber
\end{align}

With the help of Eq.\eqref{LLD} in appendix, we finally obtain the analytical form of the $\rho_0$ polarization functions Eq.\eqref{rho00} and Eq.\eqref{rho01} (defined in Eq.\eqref{zxbiaoxiang}). The analytical form of the $\rho_+$ polarization functions Eq.\eqref{rho11}, Eq.\eqref{rho12} and Eq.\eqref{rho13} (defined in Eq.\eqref{biaoxiang}) can be obtained by Eq.\eqref{Schw} and Eq.\eqref{LLF} in appendix.

Note that, in Schwinger scheme, the particle propagator in magnetic field can also be written in the proper-time form~\cite{miransky2015quantum},
\begin{eqnarray}
	&\textstyle\tilde{S}_f(p_\perp,p_\parallel)=\int_{0}^{\infty}ds {\text {Exp}}\left[-s\left(m_q^2+p_\parallel^2+p_\perp^2\frac{tanh(|Q_fB|s)}{|Q_fBs|}\right)\right]\nonumber\\	&\textstyle\times\left\{(m_q-p_\parallel\cdot\gamma_\parallel)\left(1+is_f\gamma^1\gamma^2tanh(|Q_fB|s)\right)-\frac{p_\perp\cdot\gamma_\perp}{cosh^2(|Q_fB|s)}\right\},\nonumber
\end{eqnarray}
which is equivalent to the Landau level form. The above derivations and proof in Section A and B do not depend on different form of particle parpagators. Different schemes eventually give same results.

\section{NUMERICAL RESULTS}
\label{numerical}
The $SU(2)$ NJL model is a non-renormalizable theory. Therefore, we need to perform regularization. The discrete Landau levels in the magnetic field cause anisotropy in momentum space. To ensure causality in this anisotropic system, we use the Pauli-Villars regularization scheme~\cite{mao2016inverse,mao2017effect}. In this scheme, the quark momentum and Landau level run formally from zero to infinity, and the divergence is removed by the cancellation among the subtraction terms. One introduces the regularized quark masses $m_i=\sqrt{m^2_q+a_i\Lambda^2}$ for $i=0,1,\cdots, N$, and replaces $m^2_q$ in the quark energy $E_f=\sqrt{q^2_3+2n|Q_fB|+m^2_q}$ by $m_i^2$. And the summation and integration in gap equation and pole equations are changed as%
\begin{eqnarray}
&&\sum_n\int \frac{dq_3}{2\pi} Function(E_f)\ \ \ \longrightarrow \nonumber\\
&&\sum_n\int \frac{dq_3}{2\pi} \sum_{i=0}^N \left[ c_i \times Function(E_f^i) \right], \nonumber
\end{eqnarray}
with $E_f^i=\sqrt{q^2_3+2n|Q_fB|+m^2_i}$. The parameters $N=3$, $a_1=1, c_1=-3$, $a_2=2, c_2=3$, $a_3=3, c_3=-1$, are determined by constraints $a_0=0$, $c_0=1$, and $\sum_{i=0}^N c_im_i^{2L}=0$ for $L=0,1,\cdots N-1$.

The parameters of the NJL model are fixed by fitting the chiral condensate $\langle \bar{\psi}\psi \rangle = -(230 \text{ MeV})^3$, the $\pi$ and $\rho$ meson masses $m_\pi = 134 \text{ MeV}$ and $m_\rho = 775 \text{ MeV}$, and the pion decay constant $f_\pi = 93 \text{ MeV}$ in vacuum~\cite{he1998pipi,brauner2016vector}. This yields the current quark mass $m_0 = 6.28 \text{ MeV}$, the scalar channel coupling constant $G_S = 8.27 \text{ GeV}^{-2}$, the vector channel coupling constant $G_V = 3.87 \text{ GeV}^{-2}$, and the Pauli-Villars mass parameter $\Lambda = 891 \text{ MeV}$. 

In the numerical calculations, we firstly solve the effective quark mass $m_q$ based on gap equation Eq.\eqref{mqeb}. Then, through pole equation Eq.\eqref{poleeq}, the mass spectra of $\rho$ mesons are solved at finite magnetic field and temperature, with the polarization function defined in Eq.\eqref{rho00}--\eqref{rho13}. All Landau levels are considered.

\subsection{$T=0$}
\begin{figure}[hbt]
	\centering
	\includegraphics[width=8cm]{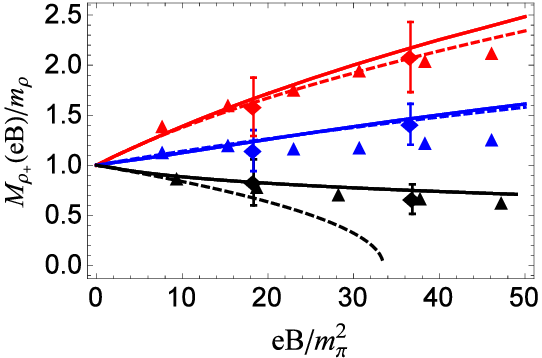}
	\caption{Masses of charged $\rho_+$ mesons as functions of magnetic field with vanishing temperature $T=0$. The red, blue, and black solid lines correspond to $\rho^{-}_+$, $\rho^{0}_+$, and $\rho^{+}_+$ mesons, respectively. The dashed line corresponds to the free point particle model. The results of LQCD calculations are depicted by triangles~\cite{luschevskaya2017determination} and diamonds~\cite{bali2018meson}, respectively. Here, $m_\rho$ is the mass of $\rho$ meson in vacuum.}
	\label{mesoneB}
\end{figure}

The masses of charged $\rho_+$ mesons with finite magnetic field and vanishing temperature are shown in Fig.\ref{mesoneB}. Both $M_{\rho^{-}_+}$ (see red solid line) and $M_{\rho^{0}_+}$ (see blue solid line) increase with magnetic field, with $M_{\rho^{-}_+}$ increasing faster than $M_{\rho^{0}_+}$. $M_{\rho^{+}_+}$ (see black solid line) decreases with increasing magnetic field, and becomes saturated at strong enough magnetic field. The results are consistent with LQCD simulations~\cite{luschevskaya2017determination,bali2018meson}. It should be mentioned that in the NJL model of different regularization schemes, the results on $M_{\rho^{-}_+}$ and $M_{\rho^{0}_+}$ are qualitatively similar. But the results of $M_{\rho^{+}_+}$ at strong magnetic field varies. For instance, with the MFIR regularization scheme, $M_{\rho^{+}_+}$ firstly decreases and then increases with magnetic field~\cite{carlomagno2022charged,Sccoccola,gomez2023charged}. 

In our current work, we consider the $\rho$ mesons as composite particles. To understand the role played by the constituent quarks, we make comparison with the results of free point particle approximation. In this approximation, the energy dispersion relation of $\rho$ mesons reads $E^2_{k_3,n,s_z}=k_3^2+\left[2n-2{\text {sign}}(Q_\rho)s_z+1\right]|Q_\rho B|+m^2_\rho$. With the lowest Landau level $n=0$ and vanishing momentum $k_3=0$, we have $E^2_{k_3=0,n=0,s_z}=\left[-2{\text {sign}}(Q_\rho)s_z+1\right]|Q_\rho B|+m^2_\rho$. For $\rho_+$ mesons, their effective masses become ${{\cal M}}_{\rho^{-}_+}=\sqrt{m^2_\rho+3|eB|}$, ${{\cal M}}_{\rho^{0}_+}=\sqrt{m^2_\rho+|eB|}$ and ${{\cal M}}_{\rho^{+}_+}=\sqrt{m^2_\rho-|eB|}$, which are plotted in dashed lines in Fig.\ref{mesoneB}. $M_{\rho^{0}_+}$ (see blue solid line) agrees well with ${{\cal M}}_{\rho^{0}_+}$ (see blue dashed line), and $M_{\rho^{-}_+}$ (see red solid line) shows some deviation from ${{\cal M}}_{\rho^{-}_+}$ (see red dashed line). However, apparent difference exists between $M_{\rho^{+}_+}$ (see black solid line) and ${{\cal M}}_{\rho^{+}_+}$ (see black dashed line), the former (latter) of which becomes saturated (an imaginary number) at strong enough magnetic field.

\begin{figure}[hbt]
	\centering
	\includegraphics[width=8cm]{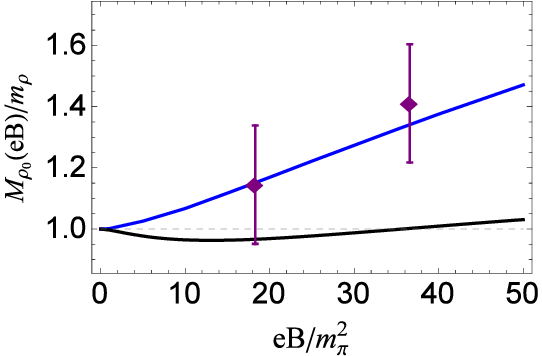}
	\caption{Masses of neutral $\rho_0$ mesons as functions of magnetic field with vanishing temperature $T=0$. The blue and black solid lines correspond to the $\rho_0^\pm$ and $\rho_0^0$ mesons, respectively. The dashed line corresponds to the free point particle model. The results of $\rho_0^\pm$ mass in LQCD calculations are shown by diamond~\cite{bali2018meson}, but LQCD did not report the results of $\rho_0^0$ mass under external magnetic field. Here, $m_\rho$ is the mass of $\rho$ meson in vacuum.}
	\label{meson0eB}
\end{figure}

The masses of the neutral $\rho_0$ mesons with finite magnetic field and vanishing temperature are shown in Fig.\ref{meson0eB}. The $\rho_0^\pm$ mesons are degenerate in the magnetic field and their mass increases with the magnetic field (see blue solid line), which is consistent with LQCD results and theoretical calculations~\cite{bali2018meson,avancini2022magnetized,larina2014rho,Sccoccola}. The mass of the $\rho^0_0$ meson is not sensitive to magnetic field, which slightly decreases and then slightly increases as magnetic field grows upto $eB=50m^2_{\pi}$ (see black solid line). This is qualitatively consistent with previous theoretical calculations~\cite{avancini2022magnetized}, but the LQCD result is not available. For neutral mesons, in free point particle approximation, they are not influenced by the external magnetic field, with constant masses ${{\cal M}}_{\rho_0^{0,\pm}}=m_\rho$ (see dashed line). When considering their constituent quarks, $M_{\rho_0^\pm }$ is enhanced by magnetic field with $M_{\rho_0^\pm } > {{\cal M}}_{\rho_0^{\pm}}$, but $M_{\rho_0^0 } \simeq {{\cal M}}_{\rho_0^{0}}$. Note that the increasing slope of $M_{\rho_0^\pm }$ in Fig.\ref{meson0eB} is similar as $M_{\rho^{0}_+}$ in Fig.\ref{mesoneB}.

When solving pole equations for $\rho$ mesons at finite magnetic field, the infrared divergence may happen in the integral term $\frac{1}{k_{0} -E_{n} +E_{n'}}$ of the polarization functions, due to the dimension reduction of their constituent quarks under external magnetic field. This leads to the multiple solutions for the meson mass at finite magnetic field. In Fig.\ref{mesoneB} and Fig.\ref{meson0eB}, we plot the mass solution of the lowest value. The solutions with higher value exist, as shown in Fig.\ref{rho1}, Fig.\ref{rho0} and Fig.\ref{firstpole}. Similar phenomena are observed in the studies of other mesons under external magnetic field, such as $\pi_{0}$, $\pi_{\pm}$, $K_{0}$, $K_{\pm}$, $\eta$, $\eta'$ and $\phi$ mesons~\cite{sheng2021pole,sheng2022impacts,Li:2025wqb,Mei:2022dkd,mao2017effect,shengxinli,ghosh2020thermo,mao2019pions,Mei:2026xlj}.


\subsection{$T \neq 0$}

\begin{figure}[hbt]
	\centering
	\includegraphics[width=8cm]{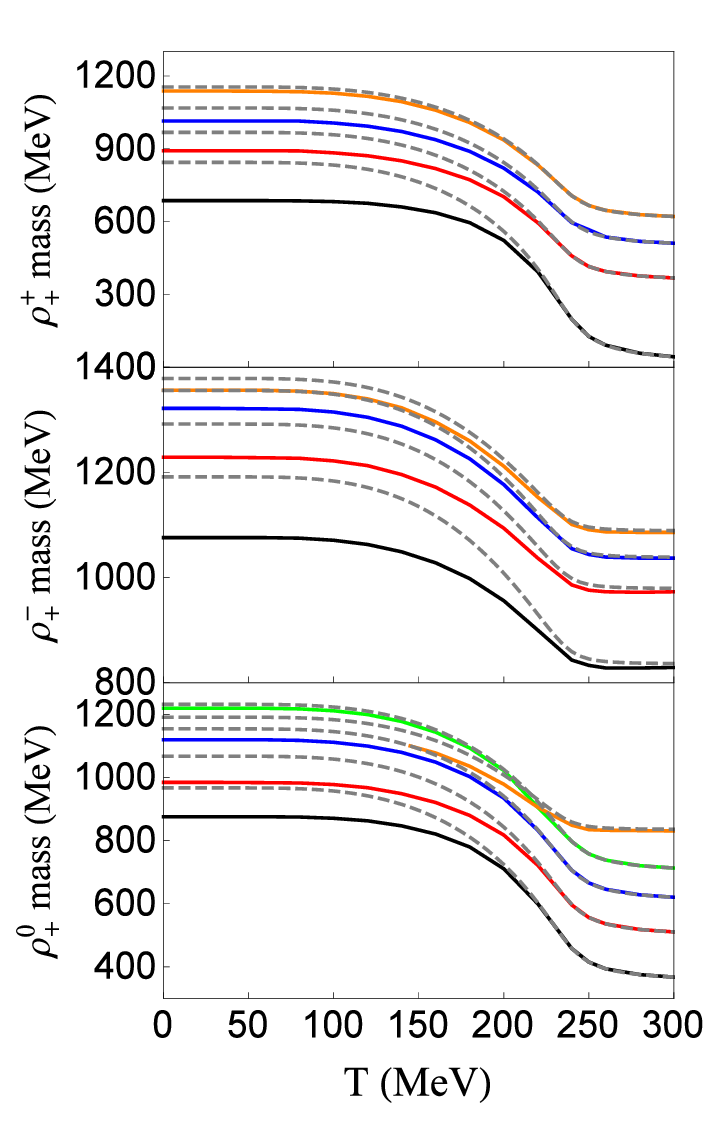}
	\caption{The lowest four/five solutions of $\rho_+$ masses as functions of temperature with finite magnetic field $eB=10m^2_{\pi}$, plotted in colored solid lines. The dashed lines show the mass sum of the constituent quarks $m_u^{(n)}+m_d^{(n')}=\sqrt{m_q^2+2n|Q_u B|}+\sqrt{m_q^2+2n'|Q_d B|}$, with integer $n$ and $n'$.}
	\label{rho1}
\end{figure}

Figure \ref{rho1} plots the results of lowest four/five solutions of $\rho_+$ masses, $M_{\rho_+^+}^{i=0,1,2,3}$ (upper panel), $M_{\rho_+^-}^{i=0,1,2,3}$ (middle panel) and $M_{\rho_+^0}^{i=0,1,2,3,4}$ (lower panel), as functions of temperature with finite magnetic field $eB=10m^2_{\pi}$. The mass sum of the constituent quarks $m_u^{(n)}+m_d^{(n')}=\sqrt{m_q^2+2n|Q_u B|}+\sqrt{m_q^2+2n'|Q_d B|}$ are also plotted in dashed lines. In Fig.\ref{rho1} upper panel, $M_{\rho_+^+}^{i=0,1,2,3}$ decreases with temperature. It will approach the mass sum of constituent quarks $m_u^{(n)}+m_d^{(n')}$ at high temperature with $M_{\rho_+^+}^{i=0} \simeq m_u^{(n=0)}+m_d^{(n'=0)}$, $M_{\rho_+^+}^{i=1} \simeq m_u^{(n=0)}+m_d^{(n'=1)}$, $M_{\rho_+^+}^{i=2} \simeq m_u^{(n=0)}+m_d^{(n'=2)}$, and $M_{\rho_+^+}^{i=3} \simeq m_u^{(n=0)}+m_d^{(n'=3)}$. In Fig.\ref{rho1} middle panel, $M_{\rho_+^-}^{i=0,1,2,3}$ decreases with temperature. It will approach the mass sum of constituent quarks $m_u^{(n)}+m_d^{(n')}$ at high temperature with $M_{\rho_+^-}^{i=0} \simeq m_u^{(n=1)}+m_d^{(n'=1)}$, $M_{\rho_+^-}^{i=1} \simeq m_u^{(n=1)}+m_d^{(n'=2)}$,  $M_{\rho_+^-}^{i=2} \simeq m_u^{(n=2)}+m_d^{(n'=1)}$ and $M_{\rho_+^-}^{i=3} \simeq m_u^{(n=1)}+m_d^{(n'=3)}$. In Fig.\ref{rho1} lower panel, $M_{\rho_+^0}^{i=0,1,2,3,4}$ decreases with temperature. It will approach the mass sum of constituent quarks $m_u^{(n)}+m_d^{(n')}$ at high temperature with $M_{\rho_+^0}^{i=0} \simeq m_u^{(n=0)}+m_d^{(n'=1)}$, $M_{\rho_+^0}^{i=1} \simeq m_u^{(n=0)}+m_d^{(n'=2)}$, $M_{\rho_+^0}^{i=2} \simeq m_u^{(n=0)}+m_d^{(n'=3)}$, $M_{\rho_+^0}^{i=3} \simeq m_u^{(n=1)}+m_d^{(n'=1)}$, and $M_{\rho_+^0}^{i=4} \simeq m_u^{(n=0)}+m_d^{(n'=4)}$. $M_{\rho_+^0}^{i=3}$ only appears at high temperature. $M_{\rho_+^0}^{i=3}$ and $M_{\rho_+^0}^{i=4}$ cross with each other at high temperature, which is associated with the crossing of the constituent quark mass sum $m_u^{(n=1)}+m_d^{(n'=1)}$ and $m_u^{(n=0)}+m_d^{(n'=4)}$.

\begin{figure}[hbt]
	\centering
	\includegraphics[width=8cm]{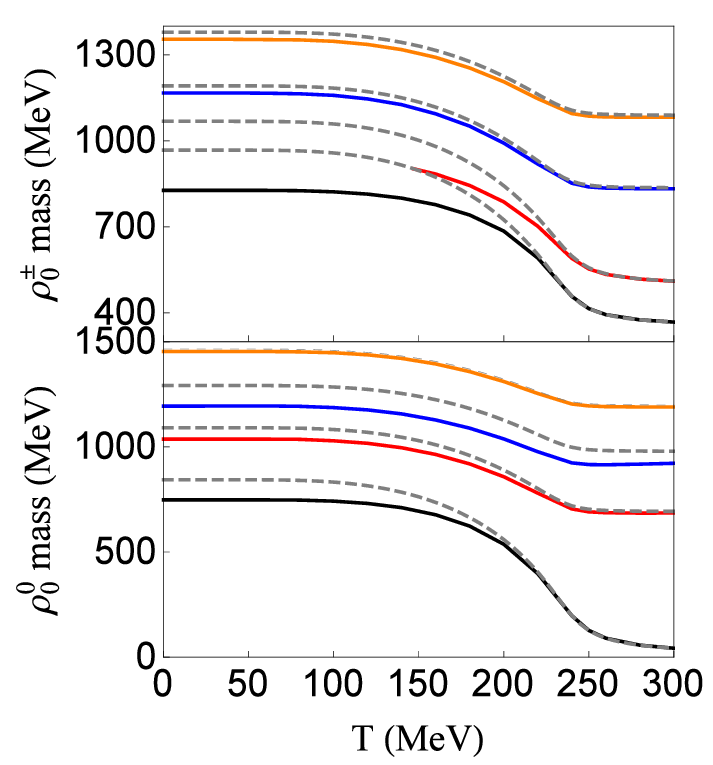}
	\caption{The lowest four solutions of $\rho_0$ masses as functions of temperature with finite magnetic field $eB=10m^2_{\pi}$, plotted in colored solid lines. The dashed lines show the mass sum of constituent quarks $m_d^{(n)}+m_d^{(n')}=\sqrt{m_q^2+2n|Q_d B|}+\sqrt{m_q^2+2n'|Q_d B|}$ (or equivalently $m_u^{(l)}+m_u^{(l')}=\sqrt{m_q^2+2l|Q_u B|}+\sqrt{m_q^2+2l'|Q_u B|}$), with integer $n$ and $n'$ (with integer $l$ and $l'$).}
	\label{rho0}
\end{figure}

Figure \ref{rho0} plots the results of lowest four solutions of $\rho_0$ masses, $M_{\rho_0^\pm}^{i=0,1,2,3}$ (upper panel) and $M_{\rho_0^0}^{i=0,1,2,3}$ (lower panel), as functions of temperature with finite magnetic field $eB=10m^2_{\pi}$. The mass sum of the constituent quarks $m_d^{(n)}+m_d^{(n')}=\sqrt{m_q^2+2n|Q_d B|}+\sqrt{m_q^2+2n'|Q_d B|}$ (or equivalently $m_u^{(l)}+m_u^{(l')}=\sqrt{m_q^2+2l|Q_u B|}+\sqrt{m_q^2+2l'|Q_u B|}$) are also plotted in dashed lines. In Fig.\ref{rho0} upper panel, $M_{\rho_0^\pm}^{i=0,1,2,3}$ decreases with temperature. It will approach the mass sum of constituent quarks $m_d^{(n)}+m_d^{(n')}$ (or equivalently $m_u^{(l)}+m_u^{(l')}$) at high temperature with $M_{\rho_0^\pm}^{i=0} \simeq m_d^{(n=0)}+m_d^{(n'=1)}$, $M_{\rho_0^\pm}^{i=1} \simeq m_u^{(l=0)}+m_u^{(l'=1)}$, $M_{\rho_0^\pm}^{i=2} \simeq m_d^{(n=1)}+m_d^{(n'=2)}$, and $M_{\rho_0^\pm}^{i=3} \simeq m_d^{(n=2)}+m_d^{(n'=3)}$. Note that $M_{\rho_0^\pm}^{i=1}$ appears at high temperature region. In Fig.\ref{rho0} lower panel, $M_{\rho_0^0}^{i=0,1,2,3}$ decreases with temperature. It will approach the mass sum of constituent quarks $m_d^{(n)}+m_d^{(n')}$ (or equivalently $m_u^{(l)}+m_u^{(l')}$) at high temperature with $M_{\rho_0^0}^{i=0} \simeq 2m_d^{(n=0)}=2m_u^{(l=0)}$, $M_{\rho_0^0}^{i=1} \simeq 2m_d^{(n=1)}$, $M_{\rho_0^0}^{i=2} \simeq 2m_d^{(n=2)}=2m_u^{(l=1)}$, and $M_{\rho_0^3}^{i=3} \simeq 2m_d^{(n=3)}$. No crossing is observed in $M_{\rho_0^\pm}^{i=0,1,2,3}$ or $M_{\rho_0^0}^{i=0,1,2,3}$, which is different from the situation of $\rho_+$ mesons.

\begin{figure}[hbt]
	\centering
\includegraphics[width=8cm]{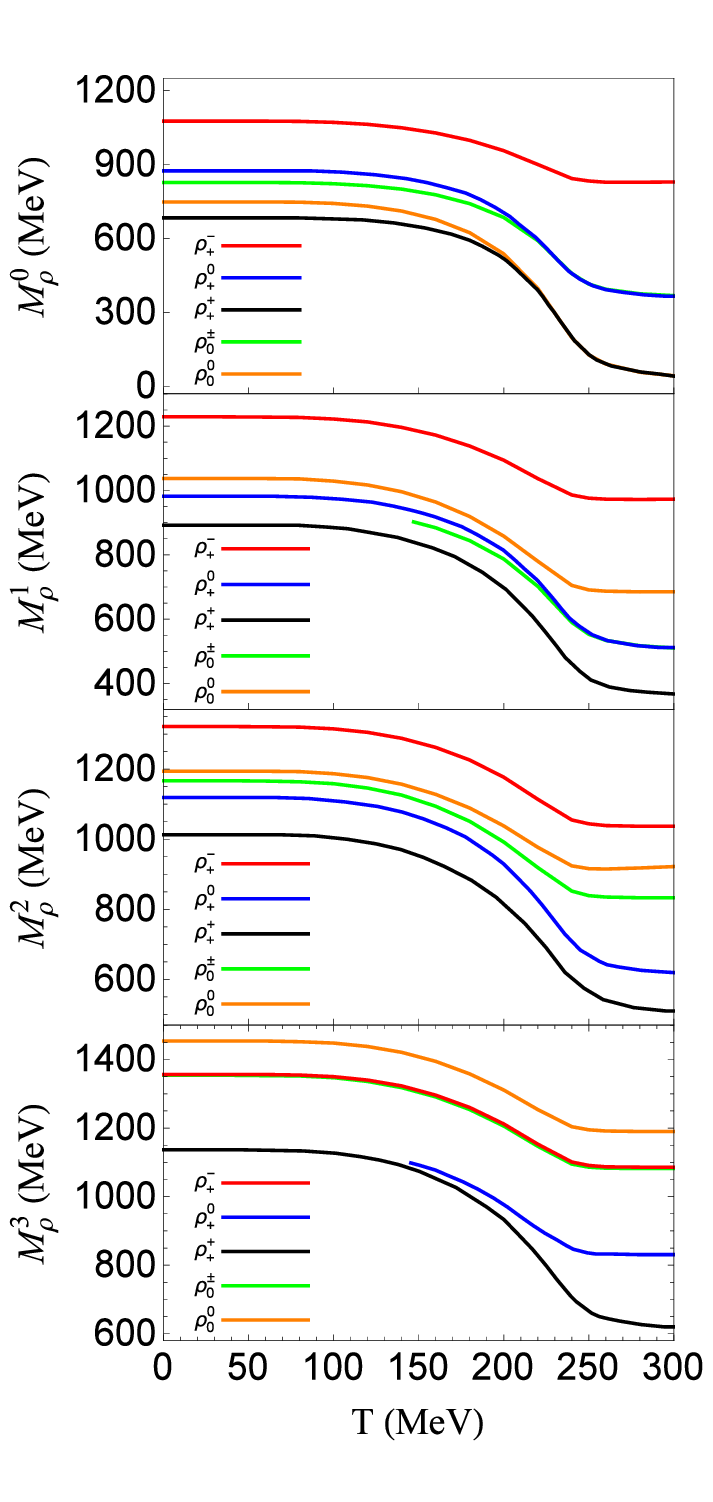}
	\caption{The masses of $\rho$ mesons, $M^{i=0}_\rho$ (upper panel), $M^{i=1}_\rho$ (second panel), $M^{i=2}_\rho$ (third panel), and $M^{i=3}_\rho$ (lower panel), as functions of temperature with fixed magnetic field $eB = 10 m_{\pi}^2$.}
	\label{firstpole}
\end{figure}

In Fig.\ref{rho1} and Fig.\ref{rho0}, we observe that at high temperature, the meson mass gradually becomes dominated by the constituent quark mass, suggesting that different vector mesons will be degenerate. Figure~\ref{firstpole} depicts the $M^{i=0}_\rho$ (upper panel), $M^{i=1}_\rho$ (second panel), $M^{i=2}_\rho$ (third panel), and $M^{i=3}_\rho$ (lower panel) as functions of temperature for $\rho$ mesons. For $M^{i=0}_\rho$, the charged $\rho_+^+$ meson becomes degenerate with the neutral $\rho_0^0$ meson at high temperature, and the charged $\rho_+^0$ meson becomes degenerate with the neutral $\rho_0^\pm$ mesons at high temperature. For $M^{i=1}_\rho$, we find $\rho_+^0$ and $\rho_0^\pm$ remain degenerate at high temperature. There exists no degeneracy in $M^{i=2}_\rho$. For $M^{i=3}_\rho$, $\rho_0^\pm$ and $\rho_+^-$ become degenerate in the whole temperature region.

\section{summary}
\label{sum}
In this paper, we perform a systematic study of the mass spectra of charged and neutral $\rho$ mesons at finite magnetic field and finite temperature within the two-flavor NJL model. To deal with the ultraviolet divergences, we adopt the Pauli-Villars regularization scheme.

The meson propagators are firstly constructed in the coordinate space using RPA method. By defining an appropriate Fourier-like transformation from coordinate space to the momentum space, which is carried by the eigenfunctions of the magnetized Proca equation for $\rho$ mesons, an algebraic formula for meson propagator is obtained in the momentum space. We conduct the analytical derivations by using Schwinger scheme and Ritus scheme of particle propagators in the magnetic field, and prove that different schemes lead to the same formula for the $\rho$ meson propagators. As far as we know, this is the first time to obtain algebraic equation of charged meson propagators in two different schemes. In fact, in our previous paper~\cite{mao2019pions,Tian}, the algebraic equations for charged $\pi$ and $K$ meson propagators have been derived in the Ritus scheme. The same algebraic formula has been obtained in the Schwinger scheme, when we prepare this paper, which is a supplementary result for our previous work~\cite{mao2019pions,Tian} and thus not listed here. In the near future, we will apply the similar method to investigate the properties of baryons under external magnetic field.

Moreover, we have numerically calculated the mass spectra of $\rho$ mesons at finite magnetic field and finite temperature. Different from the case without magnetic field, multiple solutions for $\rho$ meson mass are obtained with non-vanishing magnetic field, which happens at vanishing and finite temperature. The reason lies in the dimension reduction of their constituent quarks in magnetic fields.

At vanishing temperature, we discuss the $\rho$ meson masses $M_{\rho}$ corresponding to the lowest value solution of the pole equation. $M_{\rho^{-}_+}$, $M_{\rho^{0}_+}$ and $M_{\rho^{\pm}_0}$ increase with magnetic field. Among them, $M_{\rho^{-}_+}$ shows the fastest increasing slope, and $M_{\rho^{0}_+}$ has similar increasing slope as $M_{\rho^{\pm}_0}$. $M_{\rho^{+}_+}$ decreases with increasing magnetic field, and becomes saturated at strong enough magnetic field. $M_{\rho^0_0}$ is not sensitive to magnetic field, which slightly decreases and then slightly increases as magnetic field grows upto $eB=50m^2_{\pi}$. The results are consistent with the available LQCD simulations.

At finite temperature, we study the lowest four/five solutions of $\rho$ meson mass $M^{i=0,1,2,3,4}_{\rho}$. With fixed magnetic field, they decrease with temperature, and approach the mass sum of their constituent quarks at high temperature. The mass solution $M^{i}_{\rho}$ for different mesons $\rho_+^{0,\pm}$ and $\rho_0^{0,\pm}$ may become degenerate at finite magnetic field and temperature.

It should be mentioned that, when the meson mass exceeds the sum of its constituent quark masses, it becomes a resonant state with a finite width. In this case, the pole equation \eqref{poleeq} should be solved with a complex mass $M_m+i\Gamma_m$. Considering the width $\Gamma_m$ is much smaller than meson mass $M_m$, we still adopt the real pole equation \eqref{poleeq} in current work for simplicity~\cite{wang2018meson,zhang2016properties,liu2018neutral,li2021gauge,li2022light,li2023inverse,coppola2019neutral,xu2021effect,orlovsky2013nambu,Li:2025wqb,Mei:2022dkd}, which will not change our conclusions for meson mass spectra. In addition, we neglect the mixing between $\pi$ and $\rho$ mesons. Since $\pi-\rho$ mixing will not change the mass spectra qualitatively~\cite{bali2018meson}, it is expected that our results are robust to the $\pi-\rho$ mixing effect.

When treating mesons and baryons as composite particles, they present different properties from the free point particle approximation. Hence, the strong electromagnetic field can be used to probe the inner structure of mesons and baryons. Meanwhile, the variation of meson/baryon masses can help to identify existence of electromagnetic field in relativistic heavy-ion collisions. This issue needs further investigation in the future.\\

\noindent {\bf Acknowledgement:} Shijun Mao is supported by the National Natural Science Foundation of China under Grant No.12275204. Guoyun Shao is supported by the National Natural Science Foundation of China under Grant No.12475145 and Natural Science Basic Research Plan in Shaanxi Province of China (Program No. 2024JC-YBMS-018).\\

\section{appendix}
\label{adx}

\begin{widetext}

In the appendix, we provide some detailed derivations for polarization functions.

The integrals in Eq~\eqref{rho11jihua3}, Eq~\eqref{rho11jihua4}, Eq~\eqref{rho11jihua}, Eq~\eqref{rho11jihua1}, Eq~\eqref{rho11jihua2} fall into two classes: one contains $\alpha$ term (see Eq.\eqref{ABC} and Eq.\eqref{A2}), with identical Landau levels $g_n^{s_f}(x_1,k_2)g_n^{s_f}(y_1,k_2)$ or $I_ng_{n-1}^{s_f}(x_1,k_2)I_ng_{n-1}^{s_f}(y_1,k_2)$ and the other contains $\beta$ term (see Eq.\eqref{B1} and Eq.\eqref{B2}), with different Landau levels $I_ng_{n-1}^{s_f}(x_1,k_2)g_n^{s_f}(y_1,k_2)$ or $g_{n}^{s_f}(x_1,k_2)I_ng_{n-1}^{s_f}(y_1,k_2)$.

For the case with $\alpha$ term, the integral can be factorized according to the momenta of the constituent quarks and evaluated separately. Taking $\alpha_+^{s_u}$ as an example, we have
\begin{align}
	\label{flnn}
	&\int dp_2 e^{ip_2(x_2-y_2)}g_n^{s_u}(x_1,p_2)g_n^{s_u}(y_1,p_2)\nonumber\\
	&=\int(2^{n} {n}!\sqrt{\pi}l_u)^{-1}e^{-p^2}H_{n}\left(p+z\right)H_{n}\left(p+w\right)e^{-\frac{(x_\perp-y_\perp)^2}{4l_u^2}}e^{i\Phi_u(x_\perp,y_\perp)}dp_2\nonumber\\
	&=l_u^{-2}e^{i\Phi_u(x_\perp,y_\perp)}L_{n}\left(\frac{(x_\perp-y_\perp)^2}{2l_u^2}\right)e^{-\frac{(x_\perp-y_\perp)^2}{4l_u^2}}\nonumber\\
	&=e^{i\Phi_u(x_\perp,y_\perp)}\int2(-1)^{n}e^{-\tilde{p}_\perp^2l_u^2}L_{n}(2\tilde{p}_\perp l_u^2)e^{i\tilde{p}_\perp(x_\perp-y_\perp)}\frac{d^2\tilde{p}_\perp}{2\pi}
\end{align}
with
\begin{align}
	p=&-p_{2} l_{u} +( x_{1} +y_{1}) /2l_{u} +i( x_{2} -y_{2}) /2l_{u} ,\ \  z =( x_{1} -y_{1}) /2l_{u} -i( x_{2} -y_{2}) /2l_{u} ,\ \  w =-( x_{1} -y_{1}) /2l_{u} -i( x_{2} -y_{2}) /2l_{u}.\nonumber
\end{align}
To obtain Eq.\eqref{flnn}, we firstly perform a change of variables and reorganize the expression to isolate the Schwinger phase $\Phi_u(x_\perp,y_\perp)$. Subsequently, we apply Eq.\eqref{HHL} to integrate out the polynomial $H_n$, which gives the Laguerre polynomial in the coordinate representation. Finally, using the inverse transformation of the integral in Eq.\eqref{LJL}, we arrive at the momentum integral of the Laguerre polynomial.
\begin{align}
	\label{HHL}
	\int  dqe^{-q^{2}} H_{m}( q+z) H_{n}( q+w) &=2^{n} m!\sqrt{\pi } w^{n-m} L_{m}^{n-m}( -2zw) , n >m.\\
	\label{LJL}
	\int_{0}^{\infty}x^{\nu+1}e^{-\alpha x^2}L_n^\nu(\beta x^2)J_\nu(xy)dx&=2^{-\nu-1}\alpha^{-\nu-n-1}(\alpha-\beta)^ny^\nu e^{-\frac{y^2}{4\alpha}}L_n^\nu\left[\frac{\beta y^2}{4\alpha(\beta-\alpha)}\right].
\end{align}

For the case with the $\beta$ term, the $\beta_{\pm}^{s_f}\beta_{\pm}^{s_{f'}}$ terms can still be separated into two parts and integrated in a manner similar to the case of $\alpha$ term. But the inverse transformation of the integral in the last step requires us to combine the two $\beta_{\pm}^{s_f}\beta_{\pm}^{s_{f'}}$ terms, because the Landau levels of quarks are no longer identical. Taking $\beta_-^{s_u}\beta_+^{s_d}$ as an example, we have

\begin{align}
\label{flnn1}
	&\ \ \ \int dp_2dq_2 e^{i(p_2-q_2)(x_2-y_2)}I_{n'}g_{n'-1}^{s_u}(x_1,p_2)g_{n'}^{s_u}(y_1,p_2)g_n^{s_d}(x_1,q_2)I_ng_n^{s_d}(y_1,q_2)\nonumber\\
	&=\int(2^{n'-1} {(n'-1)}!\sqrt{\pi}l_u)^{-1}(2n')^{-1/2}I_{n'}H_{n'-1}\left(p+z_1\right)H_{n'}\left(p+w_1\right)e^{-p^2}e^{-\frac{(x_\perp-y_\perp)^2}{4l_u^2}}e^{i\Phi_u(x_\perp,y_\perp)}dp_2\nonumber\\
	&\ \ \ \ \int(2^{n-1} {(n-1)}!\sqrt{\pi}l_d)^{-1}(2n)^{-1/2}I_n H_{n}\left(q+z_2\right)H_{n-1}\left(q+w_2\right)e^{-q^2}e^{-\frac{(x_\perp-y_\perp)^2}{4l_d^2}}e^{-i\Phi_d(x_\perp,y_\perp)}dq_2\nonumber\\
	&=e^{i\Phi_+(x_\perp,y_\perp)}\frac{-(x_\perp-y_\perp)^2}{4l_ul_d}I_{n'}I_n(l_ul_d)^{-2}4(4nn')^{-1/2}L_{n'-1}^1\left(\frac{(x_\perp-y_\perp)^2}{2l_u^2}\right)e^{-\frac{(x_\perp-y_\perp)^2}{4l_u^2}}L_{n-1}^1\left(\frac{(x_\perp-y_\perp)^2}{2l_d^2}\right)e^{-\frac{(x_\perp-y_\perp)^2}{4l_d^2}}\nonumber\\
	&=-e^{i\Phi_+(x_\perp,y_\perp)}\int16\sqrt{4nn'}l_ul_d P_1 Q_1 I_n I_{n'}(-1)^{n+n'}e^{-\tilde{p}_\perp^2l_u^2}e^{-\tilde{q}_\perp^2l_d^2}L_{n'-1}^1(2\tilde{p}_\perp l_u^2)L_{n-1}^1(2\tilde{q}_\perp l_d^2)e^{i\left(\tilde{p}_\perp-\tilde{q}_\perp\right)(x_\perp-y_\perp)}\frac{d^2\tilde{p}_\perp d^2\tilde{q}_\perp}{(2\pi)^2}\nonumber\\
	&\ \ \ -e^{i\Phi_+(x_\perp,y_\perp)}\int16\sqrt{4nn'}l_ul_d P_2 Q_2 I_n I_{n'}(-1)^{n+n'}e^{-\tilde{p}_\perp^2l_u^2}e^{-\tilde{q}_\perp^2l_d^2}L_{n'-1}^1(2\tilde{p}_\perp l_u^2)L_{n-1}^1(2\tilde{q}_\perp l_d^2)e^{i\left(\tilde{p}_\perp-\tilde{q}_\perp\right)(x_\perp-y_\perp)}\frac{d^2\tilde{p}_\perp d^2\tilde{q}_\perp}{(2\pi)^2}
\end{align}
with
\begin{align}
	p&=-p_{2} l_{u} +( x_{1} +y_{1}) /2l_{u} +i( x_{2} -y_{2}) /2l_{u} ,&&z_{1} =( x_{1} -y_{1}) /2l_{u} -i( x_{2} -y_{2}) /2l_{u} ,&&&w_{1} =-( x_{1} -y_{1}) /2l_{u} -i( x_{2} -y_{2}) /2l_{u},\nonumber\\
	q&=q_{2} l_{d} +( x_{1} +y_{1}) /2l_{d} +i( x_{2} -y_{2}) /2l_{d} ,&&z_{2} =( x_{1} -y_{1}) /2l_{d} -i( x_{2} -y_{2}) /2l_{d} ,&&&w_{2} =-( x_{1} -y_{1}) /2l_{d} -i( x_{2} -y_{2}) /2l_{d}.\nonumber
\end{align}
The procedure here is identical to that used in Eq.\eqref{flnn}. During the evaluation of the integral, we isolate the Schwinger phase $\Phi_u(x_\perp,y_\perp)$ and $\Phi_d(x_\perp,y_\perp)$. It is important to note that the inverse transformation of the integral yields only the first term. The second term must be introduced by Eq.\eqref{PQ2}. Let $\phi$ and $\eta$ denote the angular coordinates of the vectors $\tilde{p}_\perp$ and $\tilde{q}_\perp$, respectively, with the components defined as $P_1 = \tilde{p}_\perp \cos(\phi)$, $P_2 = \tilde{p}_\perp \sin(\phi)$, $Q_1 = \tilde{q}_\perp \cos(\eta)$, $Q_2 = \tilde{q}_\perp \sin(\eta)$, and make use of the integral
\begin{equation}
	\label{PQ2}
	\frac{1}{2\pi}\int_{0}^{2\pi}sin(\theta)e^{izcos(\theta)}d\theta=0,
\end{equation}
we obtain the momentum $\tilde{p}_\perp\cdot\tilde{q}_\perp$ in the Schwinger scheme.

In order to derive the analytical expression for the $\rho_0$ meson polarization function Eq.\eqref{rho00} and Eq.\eqref{rho01}, the Laguerre polynomials or the generalized Laguerre polynomials must be integrated by Eq.\eqref{LLD}.
\begin{align}
	\label{LLD}
	\int_{0}^{\infty}e^{-x}x^\nu L_n^\nu(x)L_m^\nu(x)dx=&0,\ \ \left(n \neq m, \mathrm{Re}\ \nu>-1\right).\nonumber\\
	 \int_{0}^{\infty}e^{-x}x^\nu L_n^\nu(x)L_m^\nu(x)dx=&\frac{\Gamma(\nu+n+1)}{n!},\ \ \left(n = m, \mathrm{Re}\ \nu>0\right).
\end{align}
To derive the analytical expression of the $\rho_+$ meson polarization functions Eq.\eqref{rho11}, Eq.\eqref{rho12} and Eq.\eqref{rho13}, we must first perform integration over the coordinates perpendicular to the magnetic field direction,
\begin{equation}
	\label{Schw}
	\int dx_{\perp} dy_{\perp}e^{i\left(p_{\perp }-q_{\perp}\right)\left( x_{\perp } -y_{\perp }\right)} e^{i\Phi_M\left(x_\perp,y_\perp\right)} F_{n}^{*}( x,\bar{k}) F_{n}( y,\bar{k})=4\pi l_M^2(-1)^me^{-ik_{\parallel}\left(x_{\parallel}-y_{\parallel}\right)}e^{-\left(p_\perp-q_\perp\right)^2l_M^2}L_n\left(2(p_\perp-q_\perp)^2l_M^2\right).
\end{equation}
At the lowest Landau level $n=0$, we have $L_0\left(2(p_\perp-q_\perp)^2l_M^2\right)=1$. Based on Eq~\eqref{LLF}, the integration over the remaining variables ultimately leads to a Gaussian hypergeometric function,
\begin{align}
	\label{LLF}
	\int_{0}^{\infty}e^{-bx}x^\nu L_n^\nu(\lambda x)L_m^\nu(\mu x)dx=\frac{\Gamma(m+n+\nu+1)}{\Gamma(m+1)\Gamma(n+1)}\frac{(b-\lambda)^n(b-\mu)^m}{b^{m+n+\nu+1}}{}_2F_1\left[-m,-n;-m-n-\nu;\frac{b(b-\lambda-\mu)}{(b-\lambda)(b-\mu)}\right],\nonumber\\
	\left(\mathrm{Re}\ \nu>-1, \mathrm{Re}\ b>0\right).
\end{align}
Since $b(b-\lambda-\mu)/(b-\lambda)(b-\mu)=0$, the final expression takes the form of a negative binomial distribution $Z(n,n')=|eB|C^{n}_{n+n'}(2/3)^{n+1}(1/3)^{n'+1}$.

\end{widetext}

\bibliography{newreferences}

\end{document}